\documentclass[runningheads]{llncs}
\usepackage{graphicx} 
\usepackage{appendix}
\usepackage[utf8]{inputenc}
\usepackage[english]{babel}    
\usepackage{amsmath}
\usepackage{amssymb}
\usepackage{graphicx}
\usepackage{bm}
\usepackage{booktabs}
\usepackage{tabularx}
\usepackage[table,x11names]{xcolor}
\usepackage{layouts}
\usepackage{subcaption}
\usepackage[normalem]{ulem}

\definecolor{rottengreen}{rgb}{0, 0.49, 0.0}
\definecolor{dunkelgrau}{rgb}{0.4, 0.4, 0.4}

\usepackage{lineno,hyperref}
\usepackage{tcolorbox}
\modulolinenumbers[5]

\begin{document}
\title{Data assimilation in a nonlinear time-delayed dynamical system with Lagrangian optimization\thanks{T. Traverso gratefully acknowledges support from the Erasmus+ traineeship grant from the University of Genova to visit the University of Cambridge. L. Magri gratefully acknowledges support from the Royal Academy of Engineering Research Fellowships and the Hans Fischer visiting fellowship of the Technical University of Munich -- Institute for Advanced Study, funded by the German Excellence Initiative and the European Union Seventh Framework Programme under grant agreement n. 291763.}}
\titlerunning{Data assimilation in a nonlinear time-delayed dynamical system}
%
\author{Tullio Traverso\inst{1,2,3}
\and Luca Magri\inst{3,4}
}
\authorrunning{T. Traverso \and L. Magri }
%
\institute{LadHyX, Département de Mécanique, Ecole Polytechnique, \\
CNRS, 91128 Palaiseau, France\\
\email{traverso@ladhyx.polytechnique.fr}\\
\and 
Universit\`a di Genova \\
via Montallegro 1, 
 16145 Genova, Italy
 \and
Cambridge University Engineering Department, \\
Trumpington St., Cambridge, CB2 1PZ, United Kingdom \\
\email{lm547@cam.ac.uk} (L. Magri)\\
\and
Institute for Advanced Study, TU Munich (visiting fellow), \\85748 Garching, Germany\\}
\maketitle              
\begin{abstract}{
When the heat released by a flame is sufficiently in phase with the acoustic pressure, a self-excited thermoacoustic oscillation can arise. These nonlinear oscillations are one of the biggest challenges faced in the design of safe and reliable gas turbines and rocket motors~\cite{Magri2019_amr}. In the worst-case scenario, uncontrolled thermoacoustic oscillations can shake an engine apart. 
Reduced-order thermoacoustic models, which are nonlinear and time-delayed, can only qualitatively predict thermoacoustic oscillations. To make reduced-order models quantitatively predictive, we develop a data assimilation framework for state estimation. We numerically estimate the most likely nonlinear state of a Galerkin-discretized time delayed model of a horizontal Rijke tube, which is a prototypical combustor. Data assimilation is an optimal blending of observations with previous system’s state estimates (background) to produce optimal initial conditions. A cost functional is defined to measure (i) the statistical distance between the model output and the measurements from experiments; and (ii) the distance between the model’s initial conditions and the background knowledge. Its minimum corresponds to the optimal state, which is computed by Lagrangian optimization with the aid of adjoint equations. We study the influence of the number of Galerkin modes, which are the natural acoustic modes of the duct, with which the model is discretized. We show that decomposing the measured pressure signal in a finite number of modes is an effective way to enhance state estimation, especially when nonlinear modal interactions occur during the assimilation window.  This work represents the first application of data assimilation to nonlinear thermoacoustics, which opens up new possibilities for real-time calibration of reduced-order models with experimental measurements.}

\end{abstract}
\section{Nonlinear time-delayed thermoacoustic model} \label{FirstSec}
We investigate the acoustics of a resonator excited by a heat source, which is a monopole source of sound. 
The main assumptions of the reduced-order model are~\cite{Magri2019_amr}:  
(i) the acoustics evolve on top of a uniform mean flow; 
(ii) the mean-flow Mach number is negligible, therefore the acoustics are linear and no flow inhomogeneities are convected; 
(iii) the flow is isentropic except at the heat-source location;
(iv) the length of the duct is sufficiently larger than the diameter, such that the cut-on frequency is high, i.e., only longitudinal acoustics are considered;  
(v) the heat source is compact, i.e., it excites the acoustics at a specific location, $x_f$; 
(vi) the boundary conditions are ideal and  open-ended, i.e., the acoustic pressure at the ends is zero. 
Under these assumptions, the non-dimensional momentum and energy equations read, respectively~\cite{Juniper2011}
\begin{align}
\frac{\partial u}{\partial t} + \frac{\partial p}{\partial x} &= 0, \label{MomEq_ND}\\
\frac{\partial p}{\partial t} + \frac{\partial u}{\partial x}+\zeta p - \dot{\mathcal{Q}}\delta_D(x-x_f) &= 0, \label{EnEq_ND}
\end{align}
where $u$ is the acoustic velocity; 
$p$ is the acoustic pressure; 
$t$ is the time; 
$x$ is the axial coordinate of the duct; 
$\delta_D(x-x_f)$ is the Dirac delta distribution at the heat source location, $x_f$; 
$\zeta$ is the damping factor, which models the acoustic energy radiation from the boundaries and thermo-viscous losses; 
and 
$\dot{\mathcal{Q}}$ is the heat release rate (or, simply, heat release). 
The heat release, $\dot{\mathcal{Q}}$, is modelled by a nonlinear time delayed law~\cite{Orchini2016_jfm}
\begin{align}
\dot{\mathcal{Q}} \equiv \beta \textrm{Poly}(u_f(t-\tau)), \label{eq:qdotmathcal}
\end{align}
where $\tau$ is the time delay; 
$\beta$ is the strength of the heat source;  and 
$\textrm{Poly}(\tilde{u}(\tilde{t}))=a_1\tilde{u}^5(\tilde{t})+\cdots+a_5\tilde{u}(\tilde{t})$.
The time delay and strength of the heat source are the two key parameters of a reduced-order model for the flame~\cite{Crocco1969}. 
Physically, $\tau$ is the time that the heat release takes to respond to a velocity perturbation at the flame's base; while
$\beta$ provides the strength of the coupling between the heat source and the acoustics. 
Velocity, pressure, length and time are nondimensionalized as in~\cite{Juniper2011}. 
%
The set of nonlinear time-delayed partial differential equations~\eqref{MomEq_ND}-\eqref{EnEq_ND} provides a physics-based reduced-order model for the nonlinear thermoacoustic dynamics. 
Owing to the assumptions we made, the model can only qualitatively replicate the nonlinear thermoacoustic behaviour. 
In this paper, we propose a Lagrangian method to make a qualitative model quantitatively more accurate any time that reference data can be assimilated. 
Such data can come, for example, from sensors in experiments or time series from high-fidelity simulations. 
\subsection{Numerical discretization with acoustic modes} \label{DiscrThAcModel}
We use separation of variables to decouple the time and spatial dependencies of the solution. 
The spatial basis on to which the solution is projected consists of the natural acoustic modes.
When decomposed on the natural acoustic eigenfunctions, the acoustic velocity and pressure read, respectively 
\begin{align}
&u(x,t) = \sum_{j=1}^{N_{m}} \eta_j(t)\cos(j\pi x), \label{u_discr} \\
&p(x,t) = \sum_{j=1}^{N_{m}} \left(\frac{\dot{\eta}_j(t)}{j\pi}\right)\sin(j\pi x), \label{p_discr}
\end{align}
where the relationship between $\eta_j$ and $\dot{\eta}_j$ has yet to be found and $N_m$ is the number of acoustic modes considered. 
This discretization is sometimes known as the Galerkin discretization~\cite{Zinn1971}. 
The state of the system is provided by the amplitudes of the Galerkin modes that
represent velocity, $\eta_j$, and those that represent pressure, $\dot{\eta}_j/(j\pi)$. The damping has modal components, $\zeta_j=C_1j^2+C_2\sqrt{j}$, where $C_1$ and $C_2$ are damping coefficients~\cite{Landau1987,Matveev2003,Balasubramanian2008a,Juniper2011,Magri2013}. 
In vector notation, $\boldsymbol{\eta} \equiv (\eta_1,\cdots , \eta_N)^{\textrm{T}}$ and $\boldsymbol{\dot{\eta}} \equiv (\dot{\eta}_1/\pi, \cdots, \dot{\eta}_N/(N\pi))^{\textrm{T}}$.
The state vector of the discretized system is the column vector $\textbf{x} \equiv (\boldsymbol{\eta}; \boldsymbol{\dot{\eta}})$.
%
%
%
The governing equations of the Galerkin modes are found by substituting \eqref{u_discr}-\eqref{p_discr} into~\eqref{MomEq_ND}-\eqref{eq:qdotmathcal}. Equation~\eqref{EnEq_ND} is then multiplied by $\sin(k\pi x)$ and integrated over the domain $x = [0, 1]$ (projection). 
In so doing, the spatial dependency is removed and the Galerkin amplitudes are governed by a set of nonlinear time-delayed differential equations 
\begin{align}
\textrm{F}_{1j} &\equiv \frac{d}{dt} \left(\eta_j\right) - j\pi \left(\frac{\dot{\eta}_j}{j\pi}\right) = 0&t>0,\label{F1}\\
\textrm{F}_{2j} &\equiv \frac{d}{dt}\left(\frac{\dot{\eta}_j}{j\pi}\right) + j\pi \eta_j + \zeta_j\left(\frac{\dot{\eta}_j}{j \pi}\right) = 0 &t \in [0,\tau), \label{F2:1}  \\ 
\textrm{F}_{2j} &\equiv \frac{d}{dt}\left(\frac{\dot{\eta}_j}{j\pi}\right) + j\pi \eta_j + \zeta_j\left(\frac{\dot{\eta}_j}{j \pi}\right) + 2 s_j \beta \ \textrm{Poly}(u_f(t-\tau)) = 0 & t \in [\tau,T], \label{F2:2} 
\end{align}
where $u_f(t-\tau) = \sum_{j=1}^{N_{m}} \eta_j(t-\tau)c_j$; 
$s_j\equiv \sin(j\pi x_f)$ and $c_j\equiv \cos(j\pi x_f)$. 
The labels $\textrm{F}_{\bullet}$ are introduced for the definition of the Lagrangian (Sec.~\ref{sec:calcuksfeinfgr}). 
%
%
%
Because the Galerkin expansions (\ref{u_discr})-(\ref{p_discr}) are truncated at the $N_m$-th mode, we obtain a  reduced-order model of the original thermoacoustic system  (\ref{MomEq_ND})-(\ref{EnEq_ND}) with $2N_m$ degrees of freedom (\ref{F1})-(\ref{F2:2}). The reduced-order model is physically a set of $2N_m$ time-delayed oscillators, which are nonlinearly coupled through the heat release term. In the following sections, we employ 4D-Var data assimilation to improve the accuracy of such a reduced-order model\footnote{Although different from our study, it is worth mentioning that a study that combined 4D-Var data assimilation with reduced-order models of the Navier-Stokes equations based on proper orthogonal decomposition can be found in~\cite{Du2013}.}.
\section{Data assimilation as a constrained optimization problem}\label{}
\begin{figure}[!htb] 
  \centering
  \includegraphics[width=0.8\textwidth]{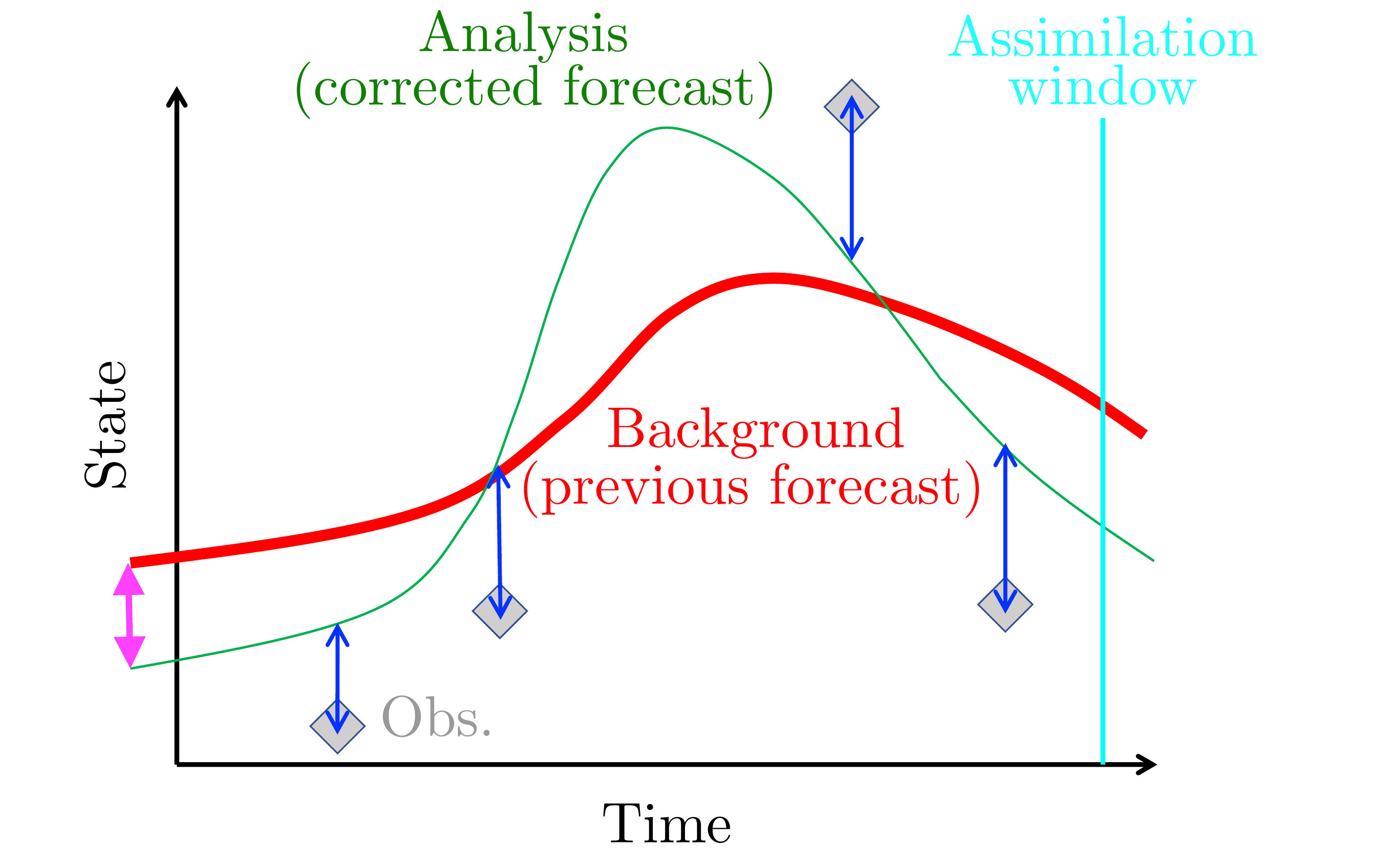}
        \caption{\small 
        Pictorial representation of data assimilation. The background error, $J_{bg}$, is proportional to the length of the magenta thick arrow, while the observation error, $J_{obs}$, is proportional to the sum of the blue thin arrows. The vertical cyan line marks the end of the assimilation window, after which the forecast begins.
        } 
        \label{PaintImage}
\end{figure}

%
%
%
%
%
The ingredients of data assimilation are 
(i) a reduced-order model to predict the amplitude of thermoacoustic oscillations, which provides the so-called background state vector $\mathbf{x}^{bg}$ at any time, $t$ (red thick line in Fig. \ref{PaintImage});
(ii) data from external observations, $\mathbf{y}^i$, which is available from high-fidelity simulations or experiments at times $t^i$ (grey diamonds in Fig. \ref{PaintImage}); and
(iii) an educated guess on the parameters of the system, which originates from past experience.
The uncertainties on the background solution and observations are here assumed normal and unbiased. 
$\mathbf{B}$ and $\mathbf{R}$ are the background and observation covariance matrices, respectively, which need to be prescribed. For simplicity, we assume that $\mathbf{B}$ and $\mathbf{R}$ are diagonal with variances $B$ and $R$ (i.e., errors are statistically independent).
A cost functional is defined to measure the statistical distance between the background predictions and the evidence from observations.
First, we want the state of the system to be as close as possible to the observations. 
Second, we do not want the improved solution to be ``too far'' away from the background solution. 
This is because we trust that our reduced-order model provides a reasonable solution. 
Mathematically, these two requirements can be met, respectively, by minimizing the following cost functional 
\begin{align}
 J
   &= \underbrace{\frac{1}{2}|| \textbf{x}_0 - \textbf{x}_0^{bg} ||_{\textrm{\textbf{B}}}^2}_{J_{bg}} + \underbrace{\frac{1}{2}\sum_{i=1}^{N_{obs}} ||\mathbf{M}\textbf{x}^{i} - \textbf{y}^{i}||_{\textrm{\textbf{R}}}^2}_{J_{obs}}\;\;\;\;\;\;\;\;\textrm{over}\;\;\;[0, T],\label{JDA} 
\end{align}
where $N_{obs}$ is the number of observations over the assimilation window $[0, T]$.
$\mathbf{M}$ is a linear measurement operator, which maps the state space to the observable space (see Eqs.~\eqref{u_discr}-\eqref{p_discr}). 
Moreover, 
$|| \textbf{x}_0 - \textbf{x}_0^{bg} ||_{\textrm{\textbf{B}}}^2 \equiv \left( \textbf{x}_0 - \textbf{x}_0^{bg} \right)^\textrm{T}\textrm{\textbf{B}}^{-1}\left( \textbf{x}_0 - \textbf{x}_0^{bg} \right)$. 
Likewise, $|| \mathbf{M}\mathbf{x}^i - \textbf{y}^i||_{\textrm{\textbf{R}}}^2 \equiv \left( \mathbf{M}\mathbf{x}^i - \textbf{y}^i \right)^\textrm{T}\textrm{\textbf{R}}^{-1}\left( \mathbf{M}\mathbf{x}^i - \textbf{y}^i \right)$. 
%
%

%
The objective of state estimation is to improve the prediction of the state, $\textbf{x}$, over the interval $[0, T]$, by reinitializing the background initial conditions, $\textbf{x}_{0}^{bg}$, with optimal initial conditions. These optimal initial conditions are called analysis initial conditions, $\textbf{x}_0^{analysis}$, which are the minimizers of the cost functional~\eqref{JDA}. 
We start from a background knowledge of the model's initial conditions, $\textbf{x}_0^{bg}$, which is provided by the reduced-order model when data is not assimilated. By integrating the system from $\textbf{x}_0^{bg}$, we obtain the red trajectory in Fig. \ref{PaintImage}, $\textbf{x}^{bg}(t)$. The set of observations is assumed to be distributed over an assimilation window at some  time instants. 
Pictorially, the analysis trajectory corresponds to the green thin line in Fig. \ref{PaintImage}, which is the minimal statistical distance between the background initial condition (magenta thick arrow) and observations (blue thin arrows). This algorithm is known as 4D-Var in weather forecasting~\cite{Blayo2014}. 
State estimation enables the adaptive update of reduced-order thermoacoustic models whenever data is available. 

\section{Data assimilation for nonlinear thermoacoustic dynamics}
We propose a set of cost functionals to perform data assimilation with the thermoacoustic model introduced in Sec.~\ref{FirstSec}. We also introduce the formalism to perform Lagrangian optimization, in which adjoint equations enable the efficient calculation of the gradients of the thermoacoustic cost functionals with respect to the initial state.
%
%
\subsection{Thermoacoustic cost functionals} \label{Thermoac_J}
Crucial to data assimilation  is the definition of the cost functionals $J_{bg}$ and $J_{obs}$. 
Three physical thermoacoustic cost functionals are proposed and compared to reproduce different scenarios.
For the background error  
\begin{align}
J_{bg}^{a} &=  \frac{1}{2B} \left( p(0)-p(0)_{bg}\right)^2 \  , \label{Jbg_a2}\\
J_{bg}^{b} &= \frac{1}{2B}\sum_{j=1}^{N_{m}} \bigg\{ \left[
\left(\frac{\dot{\eta}_{j0}}{j\pi}\right) - \left(\frac{\dot{\eta}_{j0}}{j\pi}\right)_{bg}\right]\sin(j\pi x_m) \bigg\}^2, \ \label{Jbg_b2} \\
J_{bg}^{c} &= \frac{1}{2B}\sum_{j=1}^{N_{m}} \left[
\left(\frac{\dot{\eta}_{j0}}{j\pi}\right) - \left(\frac{\dot{\eta}_{j0}}{j\pi}\right)_{bg}\right]^2 + \frac{1}{2B}\sum_{j=1}^{N_{m}}   \left[\eta_{j0} - \eta_{j0,bg}\right]^2. \label{Jbg_c2}
\end{align}
For the observation error 
\begin{align}
J_{obs}^{a} &= \sum_{i=1}^{N_{obs}}J_{obs,i}^{a} = \frac{1}{2R}\sum_{i=1}^{N_{obs}} \left( p(t_{obs}^{i})-p^{(i)}_{obs}\right)^2, \hspace{1cm}  \label{Jobs_glob3} \\
J_{obs}^{b} &= \sum_{i=1}^{N_{obs}}J_{obs,i}^{b} =  \frac{1}{2R}\sum_{i=1}^{N_{obs}}\sum_{j=1}^{N_{m}} \bigg\{ \left[
\left(\frac{\dot{\eta}_j(t_{obs}^{i})}{j\pi}\right) - \left(\frac{\dot{\eta}_j}{j\pi}\right)^{(i)}_{obs}\right]\sin(j\pi x_m) \bigg\}^2 , \hspace{1cm}  \label{Jobs_mod3}
\end{align} 
where $x_m$ is the location where the measurement is taken and $t_{obs}^{i}$ is the instant at which the $i-$th observation is assimilated. 
On the one hand, by using $J_{bg}^{a}$ and $J_{obs}^{a}$ the analysis solution is optimized against the background pressure at $t=0$ and the measured pressure at $t=t_{obs}^i$, $i=1,\dots,N_{obs}$. Physically, this means that the acoustic pressure is the model output, $p(0)_{bg}$, and the observable from the sensors, $p_{obs}^i$. 
On the other hand, $J_{bg}^{b}$ and $J_{obs}^{b}$ constrain the pressure modes. Physically, this means that every pressure mode provided by the background solution is a model output, $\left(\dot{\eta}_{j0}/(j\pi)\right)_{bg}$, and it is assumed that the modes of the acoustic pressure,$\left(\dot{\eta}_j/(j\pi)\right)^{(i)}_{obs}$, can be calculated from the sensors on the fly. For the background cost functional, we also defined $J_{bg}^c$ as a norm of the modes, which does not have a  corresponding observation cost functional because the spatial dependency is not explicit. 
%
%

%
To attain a minimum of $J$, a necessary condition is that the gradient vanishes, i.e., 
\begin{align}
\nabla_{{\textbf{x}_0}}(J) = \nabla_{{\textbf{x}_0}}(J_{bg}) +  \sum_{i=1}^{N_{obs}} \nabla_{{\textbf{x}_0}}(J_{obs,i})=0,  \label{TotObsGrad2}
\end{align}
where $\nabla_{{\textbf{x}_0}}$ is the gradient with respect to the initial conditions. 
There exists $\textbf{x}_0$ such that $\nabla_{{\textbf{x}_0}}(J) = 0$ because of the convexity of the cost functionals in the neighbourhood of a local extremum. 
To compute $\nabla_{{\textbf{x}_0}}(J_{bg})$ and $\nabla_{{\textbf{x}_0}}(J_{obs,i})$, we use calculus of variation (Sec.~\ref{sec:calcuksfeinfgr}). 
The Lagrange multipliers, also known as adjoint, or dual, or co-state variables (Sec.~\ref{AdjEq}), provide the gradient information with respect to the initial state. 
%
%
%


\subsection{Lagrangian of the thermoacoustic system}\label{sec:calcuksfeinfgr}
The governing equations and their initial conditions are rewritten in the form of constraints, $\textrm{F}$, which hold over time intervals, while $\textrm{G}$ are the constraints that hold for a specific time only, i.e., $t=t_0$. Together with equations (\ref{F1})-(\ref{F2:2}) and by defining the auxiliary variable $\bar{\eta}(t)\equiv u_f(t-\tau)$, they read 
\begin{align}
\textrm{F}_{3\ } &\equiv \bar{\eta}(t) = 0,   & t \in [0,\tau) \label{eqF3:1}\\
\textrm{F}_{3\ } &\equiv \bar{\eta}(t) - u_f(t-\tau) = 0,   &t \in [\tau,T]. \label{eqF3:2} 
\end{align}
The constraints for the initial conditions read 
\begin{align}
\textrm{G}_{1j} &\equiv \eta_j(0) - \eta_{j0} = 0, \label{G1j} \\
\textrm{G}_{2j} &\equiv \left(\frac{\dot{\eta}_j(0)}{j\pi}\right) - \left(\frac{\dot{\eta}_{j0}}{j\pi}\right) = 0. \label{G2j} 
\end{align}
By defining an inner product 
\begin{align}
    \left[a, b\right]=\frac{1}{T}\int\limits_0^T ab \ dt
    \end{align}
where $a$ and $b$ are arbitrary functions,
the Lagrangian of the nonlinear system can be written as
\begin{align}
 \mathcal{L} \equiv J_{bg} + J_{obs,i} + \sum_{j=1}^{N_{m}} \mathcal{L}_j - \left[\bar{\xi}(t),\textrm{F}_{3}\right],
 \label{L1}
\end{align}
where each $\mathcal{L}_j$ is \\
\begin{align}
\mathcal{L}_j \equiv - \left[\frac{\xi_j}{j\pi},\textrm{F}_{1j}\right] - \left[\nu_j,\textrm{F}_{2j}\right] - b_{1j}\textrm{G}_{1j} - b_{2j}\textrm{G}_{2j}, \label{L2} 
\end{align}
where 
$\bar{\xi}$, $\xi_j/j\pi$, $\nu_j$ and $b_{\bullet j}$ are the Lagrange multipliers, or adjoint variables, of the corresponding constraints.
%
Because we wish to derive the adjoint equations for the cost functional $J_{obs,i}$, we consider the time window to be $T=t_{obs}^i$. 
\subsection{Adjoint equations} \label{AdjEq}
We briefly report the steps to derive the evolution equations of the Lagrange multipliers (adjoint equations)~\cite{Magri2019_amr}. 
First, the Lagrangian \eqref{L1} is integrated by parts to make the dependence on the direct variables explicit. Secondly, the first variation is calculated by a Fr\'echet derivative 
\begin{align}
\left[ \frac{\partial \mathcal{L}}{\partial {\mathbf{x}}},\delta {\mathbf{x}}\right] \equiv \lim \limits_{\epsilon \to 0} \frac{\mathcal{L}({\mathbf{x}}+\epsilon \delta {\mathbf{x}})-\mathcal{L}({\mathbf{x}})}{\epsilon}.  \label{Deriv}
\end{align}
%
%
%
%
Thirdly, the derivatives of \eqref{L1} are taken with respect to the initial condition of each variable of the system, $\partial \mathcal{L}/\partial \textbf{x}_0$. These expressions will be used later to compute the gradient.
Finally, to find the set of Lagrange multipliers that characterizes an extremum of the Lagrangian, $\mathcal{L}$, variations with respect to $\delta \textbf{x}$ are set to zero.
The adjoint equations and their initial conditions are derived by setting variations of $\delta\eta_j$, $\delta\left(\dot{\eta}_j/(j\pi) \right)$ and $\delta\bar{\eta}$ to zero over $[0,T]$.  
\subsection{Gradient-based optimization}\label{sec:optloop}
The optimization loop consists of the following steps:
\begin{itemize}
\item[1)]Integrate the system forward from $t=0$ to $t=T$ from an initial state $\textbf{x}_0$;
\item[2)]Initialize the adjoint variables;
\item[3)]Evolve the adjoint variables backward from $t=T$ to $t=0$;
\item[4)]Evaluate the gradient using the adjoint variables at $t=0$.
\end{itemize}
Once the gradient is numerically computed, the cost functional can be minimized via a gradient based optimization loop. 
The conjugate gradient~\cite{Press2007} is used to update the cost functional until the condition $\nabla_{\textbf{x}_0}(J)=0$ is attained to a relative numerical tolerance of $10^{-4}$. 
By using a gradient based approach, we find a local minimum of $J$.
We verify that there is no other local minimum by computing $J=J(\textbf{x}_0)$ in the vicinity of $\textbf{x}_0^{analysis}$.  
%
%
\section{Results
} \label{results}
We validate the data-assimilation algorithm by twin experiments: The true state solution, $\textbf{x}^{true}(t)$, is produced by perturbing the unstable fixed point at the origin of the phase space\footnote{Strong constraint 4D-Var assumes that the model is perfect and the uncertainty is only in the initial conditions, therefore, the true trajectory can be a model output.}; the background trajectory, $\textbf{x}^{bg}(t)$, is obtained by perturbing each true mode initial condition with Gaussian error with variance $B=0.005^2$; 
the $i-$th observation is produced by adding Gaussian error with variance $R=0.005^2$ to $\textbf{x}^{true}(t^i_{obs})$.
%
The outcome of twin experiments is summarized by the error plots shown in Figs. \ref{fig:JobsEff} and \ref{Fig:EffOfNobs2}. The cyan vertical line indicates the end of the assimilation
window, the red thick line is the difference between the true pressure and the background pressure, the green thin
 line is the difference between the true pressure and the analysis pressure. 
First, it is shown how the number of computed acoustic modes affects the solution of the system.
Secondly, we investigate the effects that the different cost functionals have on the analysis solution.
Finally, we discuss the effects that the number of observations have on the analysis.

The parameters we use are $\beta=1$, $\tau=0.02$, $C_1=0.05$, $C_2=0.01$ and $(a_1, a_2, a_3, a_4, a_5) = (-0.012,0.059,-0.044,-0.108,0.5)$ for the heat release term $\dot{\mathcal{Q}}$. The position of the heat source is $x_f=0.3$ and all the measurements are taken at $x_m=0.8$.
\subsection{Remarks on the thermoacoustic nonlinear dynamics} \label{RemarksOnDyn}
\begin{figure}[!htb]
  \centering
        \includegraphics[width= 0.7\textwidth]{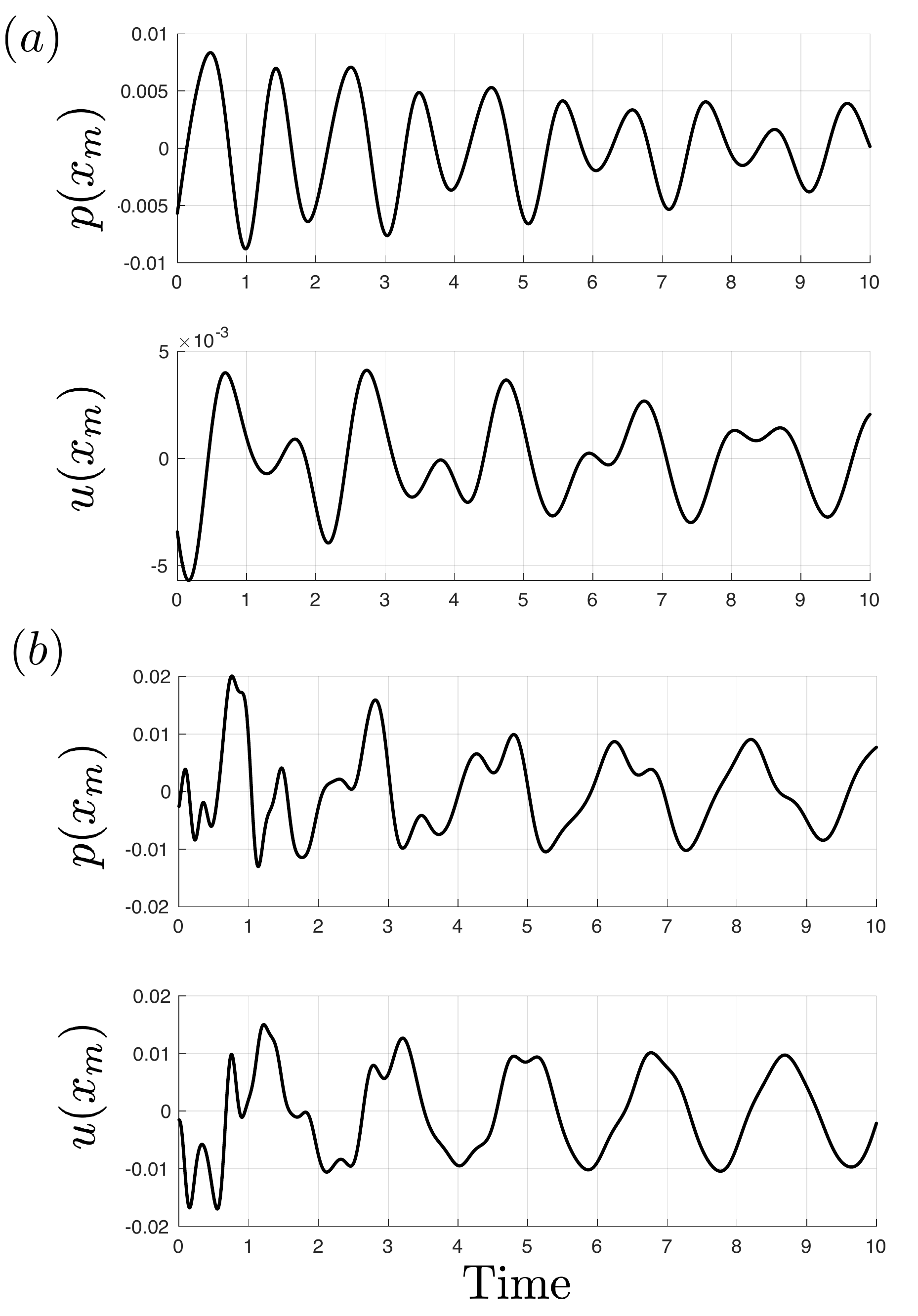}
     \caption{\small Time series of the pressure, $p$ and velocity, $u$ evaluated at $x_m=0.8$ using (a) $N_{m}=3$ and (b) $N_{m}=10$. 
     When 10 modes are computed, a transient region can be identified for $t\lesssim2$, which is characterized by irregular fluctuations due to nonlinear modal coupling. 
     } \label{TS3vs10mod}
\end{figure}

The thermoacoustic model is a set of $2N_{m}$ nonlinearly coupled oscillators, which we initialize by imposing non-equilibrium initial conditions. We compare two solutions, using $N_{m}=3$ and $N_{m}=10$ in Figs. \ref{TS3vs10mod}a and \ref{TS3vs10mod}b, respectively. Higher modes are quickly damped out, thus, after a transient where strong nonlinear modal coupling occurs, the solution obtained with $N_{m}=10$ is qualitatively similar to the solution obtained with $N_{m}=3$. During the transient, if sufficient modes are computed, the dynamics are more unpredictable because of the intricate modal interaction. The twin experiments are performed with $N_m=10$, which provide a more accurate solution. It is shown that state estimation is markedly affected depending on whether we observe the system during the transient or at regime. 
\subsubsection{Effect of the observation error} \label{J_obs_Eff}
\begin{figure}[!htb]
  \centering
  \includegraphics[width=0.7\textwidth]{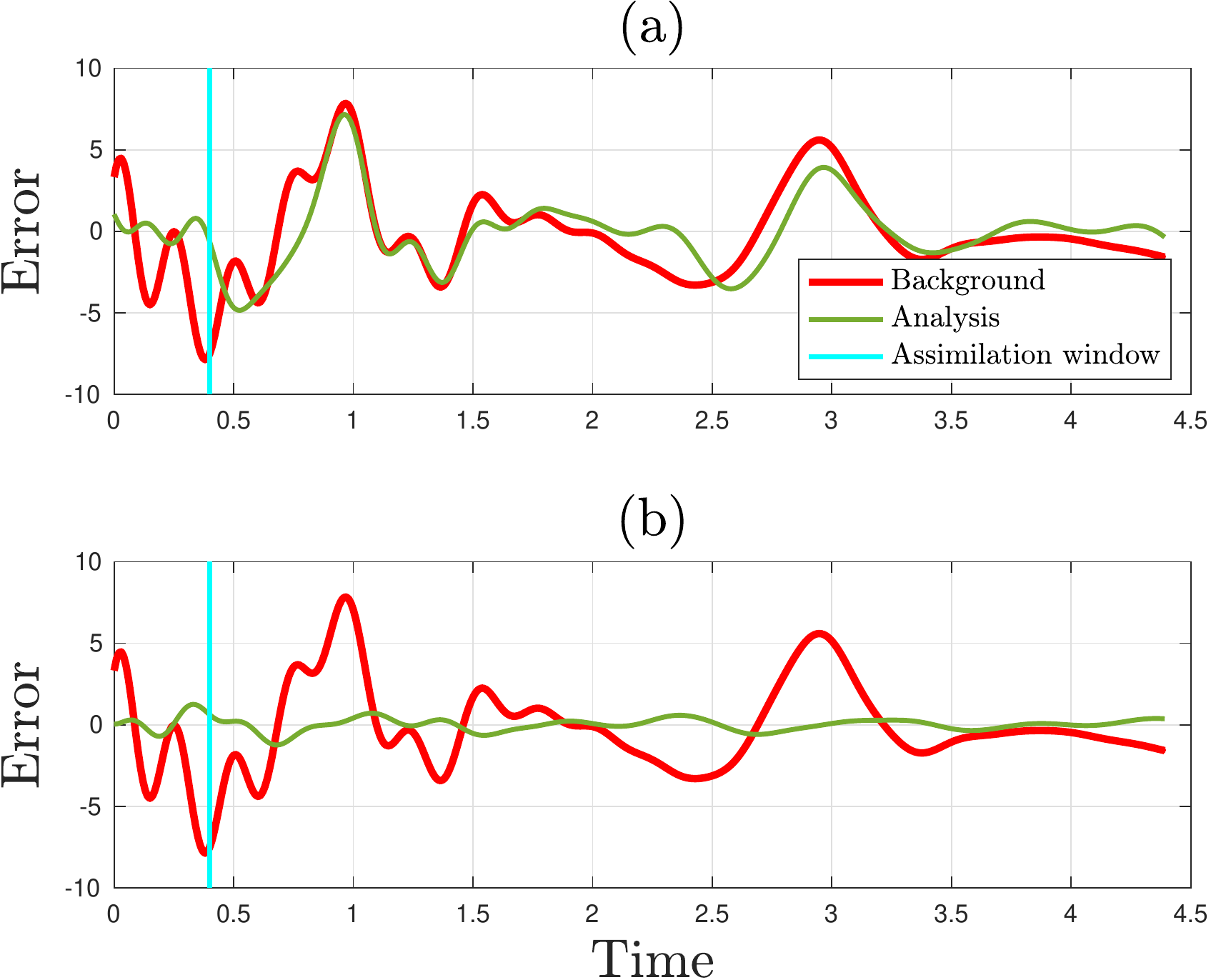}
      \caption{\small Time series of the background and analysis acoustic pressure deviation from the true state (normalized with the true acoustic pressure at $t=0$). Both twin experiments are performed with $N_{obs}=100$ (the observations are not shown). (a) The cost functional measures the (a) pressure ($J_{obs}^a$), and (b)  pressure modes ($J_{obs}^b$).} \label{fig:JobsEff}
\end{figure}
We can simulate two main scenarios, depending on the choice of $J_{obs}$ (Sec. \ref{Thermoac_J}). 
Figure \ref{fig:JobsEff}a 
is obtained using $J_{obs}^a$, i.e., by modelling observations about the pressure only. The analysis pressure error slightly deviates from zero in the assimilation window. When the forecast window starts, the analysis suddenly approaches the background again. 
Figure \ref{fig:JobsEff}b is obtained using $J_{obs}^b$, therefore the observations contain information about every pressure modes. The forecast quality is considerably enhanced. The assimilation window is $T_{as}=0.4$, thus, the observations are obtained during the transient, which lasts up to $t\approx2$, where the dynamics are more unpredictable due to nonlinear interactions between modes. As we will show in the next subsection, increasing $N_{obs}$ is not an effective strategy to improve the forecast during the transient when the pressure is observed.
\subsubsection{Effect of the background error}     
From a numerical standpoint, the background error, $J_{bg}$,  acts as an observation at $t=0$. The analysis trajectory is an optimal blending of the information from the measurements and the previous educated model output. Therefore, we emphasize that the outcome of twin experiments is improved if the cost functionals of the background knowledge and observations are consistent. In the present framework, it means that $J_{bg}^a$ should be used with $J_{obs}^a$ and $J_{bg}^b$ should be used with  $J_{obs}^b$. On the one hand, if the source of assimilated data favours the background knowledge (e.g. using $J_{bg}^b$ or $J_{bg}^c$ together with $J_{obs}^a$), the analysis trajectory will be closer to the background. On the other hand, if the source of assimilated data favours the observations (e.g. using $J_{obs}^b$ with $J_{bg}^a$), the analysis trajectory will be closer to the observations. 

\subsubsection{Effect of the number of observations}
\begin{figure}[!htb]
  \centering
          \includegraphics[width=0.7\textwidth]{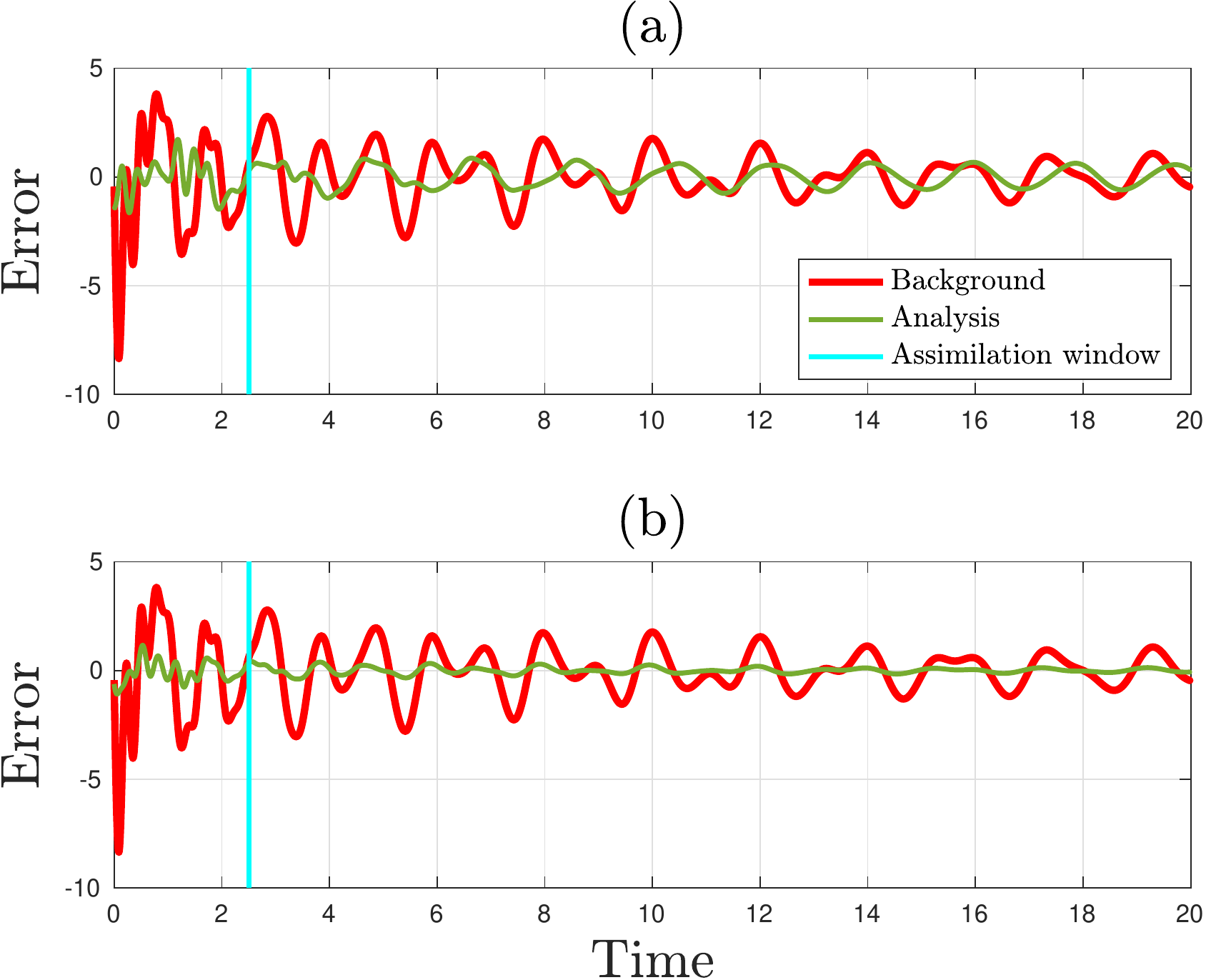} 
               \caption{\small 
     Error plots of two different twin experiments. The assimilation window is 2.5 time units and the observation error is measured using $J_{obs}^a$ in both cases. The choice of $T_{as}$ implies that the system is observed also at regime. (a) $N_{obs}=50$ and (b) $N_{obs}=250$. 
     } \label{Fig:EffOfNobs2}
\end{figure}

Generally speaking, the higher the number of observations the more the optimal solution will be similar to the true solution. This can be deduced by inspection of Figs. \ref{Fig:EffOfNobs2}a and \ref{Fig:EffOfNobs2}b. The value of $N_{obs}$ is increased from 50 to 250, over an assimilation window of 2.5 time units (the observations are not shown in the figures), resulting in a smaller error amplitude when more observations are available.
  
     %
%
However, it is possible that the increasing of $N_{obs}$ will not result in a better state estimation. 
%
This happens if we measure the pressure during the transient, that is, using $J_{obs}^a$ and $T_{as} < 2$. In the transient, the measured pressure is a combination of all modes. Therefore, the same pressure level is associated with different combinations of  modes, hence, no useful information is assimilated to help determine the state of the system by knowing only the pressure (i.e., by using $J_{obs}^a$). Under these circumstances, the forecast quality will remain poor, as shown in Fig.~\ref{fig:JobsEff}a, regardless of the number of observations. 
When the assimilation window is extended up to 2.5 time units, as shown in Fig.~\ref{Fig:EffOfNobs2}, the pressure is observed also at regime, that is in $t\in [2,2.5]$, approximately. 
In this interval, the measured pressure signal is produced chiefly by the first three modes because higher modes are damped out. 
Therefore, the measured pressure becomes more effective information about the state to assimilate. At regime, as intuitively expected, increasing the observation frequency produces a better forecast. 

We conclude that having poor information about the system's state cannot be simply balanced by increasing the number of observations. Given a number of  observations with their time distrbution, it is the synergy between an appropriate cost functional and the recognition of the type of dynamics (transient vs. regime) to be the key for a successful data assimilation.  
\section{Conclusions and future work}
Preliminary thermoacoustic design is based
on simplified and computationally cheap reduced-order models that capture the inception of thermoacoustic instabilities (linear analysis) and their saturation to finite amplitude oscillations (nonlinear analysis).
We propose a Lagrangian method to make a qualitative reduced-order model quantitatively more accurate any time that reference data can be assimilated. Such data can come, for example, from sensors
in experiments or time series from high-fidelity simulations.
To test the method we perform a series of twin experiments with the  thermoacoustic model of a resonator excited by a heat source (horizontal Rijke tube).
When sufficient modes are computed, a clear distinction emerges between a transient state and the state at regime. The former is characterized by irregular dynamics due to the interaction between all modes, while at regime the dynamics are chiefly dominated by the first three modes. We find that, at regime, it is possible to enhance the forecast by assimilating data about the pressure. As intuitively expected, the higher the number of observations, the better the forecast accuracy. While testing the effectiveness of data assimilation during the transient, we find that it is not possible to improve the forecast by measuring the pressure only. Moreover, the quality of the forecast remains poor regardless of the number of observations. Therefore, we propose a more effective cost functional, which  takes into consideration the spectral content of the measured signal to enable a successful state estimation also during the transient. 
%
%
In  state  estimation,  we  implicitly  assume  that  the  parameters are correct. However, this is rarely the case in thermoacoustics, where the parameters are uncertain and need to be optimally calibrated. Ongoing work includes simultaneous parameter and state estimation using Lagrangian optimization with state augmentation. This work opens up new possibilities for on-the-fly optimal calibration and state estimation of reduced-order models in thermoacoustics for applications in propulsion and power generation. 
\bibliographystyle{splncs04}

%
%
%
%
%
\end{document}